\documentclass[]{spie}  %>>> use for US letter paper
\usepackage{amsmath,amsfonts,amssymb}
\usepackage{graphicx}
\usepackage[colorlinks=true, allcolors=blue]{hyperref}
\usepackage{siunitx}

\title{Effects of manufacturing tolerances on the performance
of metamaterial microwave anti-reflection coatings}

\author[a]{Rustam Balafendiev}
\author[a]{Miranda Eiben}
\author[a,b]{Jon E. Gudmundsson}
\affil[a]{Science Institute, University of Iceland, 107 Reykjavik, Iceland}
\affil[b]{The Oskar Klein Centre, Department of Physics, Stockholm University,
AlbaNova, SE-10691 Stockholm, Sweden}

\authorinfo{Further author information: (Send correspondence to R.B.)\\R.B .: E-mail: rub8-at-hi-dot-is}
\begin{document} 
\maketitle

\begin{abstract}
Metamaterial anti-reflection coatings (ARC) are used in a variety of applications, including: lenses, filters, and absorbers. Typically, the design of a given ARC is done within an infinite medium approximation, which presupposes that every unit cell on the interface is identical. However, in realistic applications, the geometry of a given ARC usually has some degree of variability, be it due to the shape of the surface inherent to the application, like in a lens, or manufacturing tolerances. This variation may alter the performance of the optical element in unanticipated ways, by creating additional scattering, enabling diffractive maxima that would normally be absent and, most crucially, changing the transparency of the ARC as a function of frequency. In this work we utilize full-wave modeling of finite samples of plastic which are matched with free space using a metamaterial ARC on both of their interfaces. By adding a degree of randomness to the ARC geometry we attempt to characterize the extent to which a given variation affects the expected performance of an ARC.
\end{abstract}


\section{INTRODUCTION}
\label{sec:intro}  % \label{} allows reference to this section

A variety of transmissive optics in CMB applications, such as filters, windows and lenses, rely on metamaterial anti-reflection coatings to minimize the power lost throughout the optical path \cite{Golec:22,Nitta2018,Datta2013}. A performance of a metamaterial ARC is typically estimated via simulating a unit-cell in a frequency solver. However, this approach can not capture the impact of various cell-to-cell variations --- like machining tolerances or curved surfaces --- on the performance. In this work we propose an approach to estimating this impact using a time-domain solver and showcase its effectiveness by considering a case of random offsets in various heights of the ARC elements drawn from a Gaussian distribution.

\section{METHODS}

\begin{figure}
    \centering
    \begin{tabular}{cc}
    \includegraphics[width=0.8\linewidth]{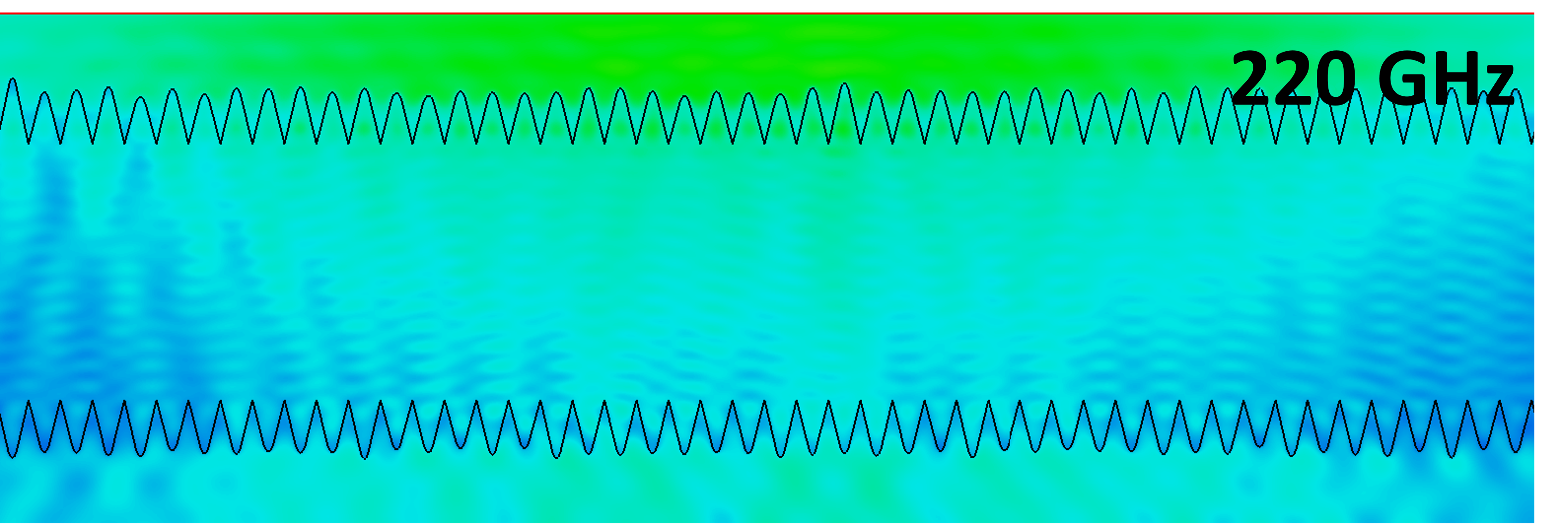}&
    \includegraphics[width=0.12\linewidth]{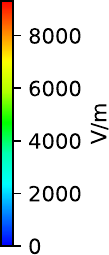}
    \end{tabular}
    \caption{An example of E-field magnitude at \SI{220}{\giga\hertz} in a nylon slab with randomized pyramid heights using $\sigma= \SI{60}{\micro\meter}$.}
    \label{fig:field}
\end{figure}

The simulations were performed using the time-domain FIT (Finite Integration Technique) solver in CST MWS. A row of 100 unit cells repeated along the $X$-direction were simulated for each case, with an 88 unit cell-wide waveguide port positioned 6$\lambda$ away being used as a Gaussian-like source. Each combination of parameters on the plots below were simulated for at least 20 different instances of geometry, with feature heights in each being randomized by a Gaussian distribution with a given standard deviation, $\sigma$ (see Fig.\ \ref{fig:field} for an example of a single iteration). The values of $\sigma$ chosen are somewhat arbitrary and present cases that vary from realistic to exaggerated. The resulting far field is then summed up using an array factor, i.e., by stacking identical far fields from sources placed one period apart, to extend the system in the $Y$-direction.

\begin{figure}
    \centering
    \includegraphics[width=0.9\linewidth]{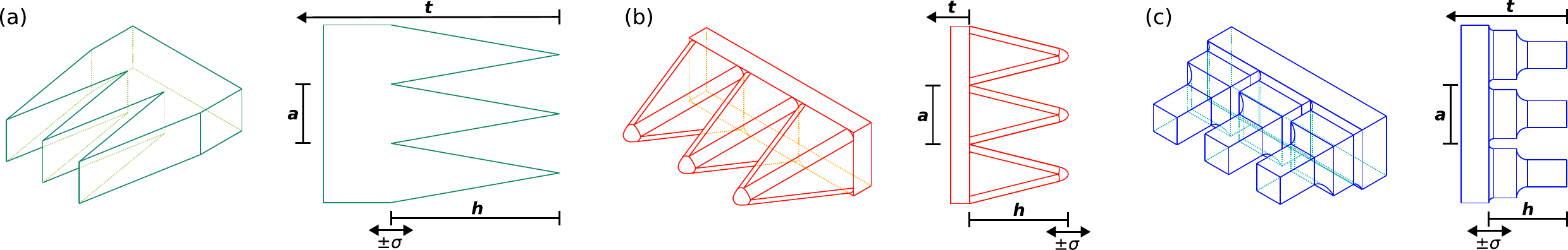}
    \caption{The geometries of the various ARCs studied. The dimensions varied by the standard deviation value sigma are indicated with a $±\sigma$ near a point that is being randomized. Cases (a) and (c) have varying depth of grooves and a fixed total sample thickness, in Case (b) the bulk nylon thickness is fixed and the height by which the pyramids extend from the plane is varied. Cases (b) and (c) are symmetric on both sides, Case (a) has a singular groove rotated by 90 degrees on the side facing the source.}
    \label{fig:geom}
\end{figure}

Three examples of ARCs were considered: perpendicular triangular grooves in HDPE investigated in Tapia et al.\cite{Tapia2018} (Fig.\ \ref{fig:geom}(a)), filleted triangular pyramids in nylon (Fig.\ \ref{fig:geom}(b)), and two-box layers in an alumina IR filter from Golec et al.\cite{Golec:22} (Fig.\ \ref{fig:geom}(c)). The frequency extents and bands for Case (a) and (b) were taken from respective papers. A single iteration took at most 6 minutes to calculate on a high-performance cluster.

\begin{table}[]
    \caption{Details on the material properties and the geometry of the three cases studied in this work. The material properties are taken from Lamb and Xue et al.\cite{Lamb1996,Xue2024}.}
    \begin{center}
    \begin{tabular}{c|c|c|c}
                 & (a) & (b) & (c)  \\
        Material & HDPE & Nylon-6 & Alumina \\
        \hline
        Index, \textit{n}  & 1.52 & 1.79 & 3.14 \\
        Loss tangent, $\tan{\delta}$ & 2.3e-4 & 200e-4 & 11.5e-4 \\
        Period, \textit{a} [mm] & 0.64 & 0.625 & 0.295 \\
        Thickness, \textit{t} [mm] & 4.36 & 5 & 5 \\
        Height, \textit{h} [mm]  & 1.82 & 1.25 & 0.388
    \end{tabular}
    \end{center}
    \label{tab:geom}
\end{table}

\section{RESULTS}
\begin{figure}
    \centering
    \includegraphics[width=1\linewidth]{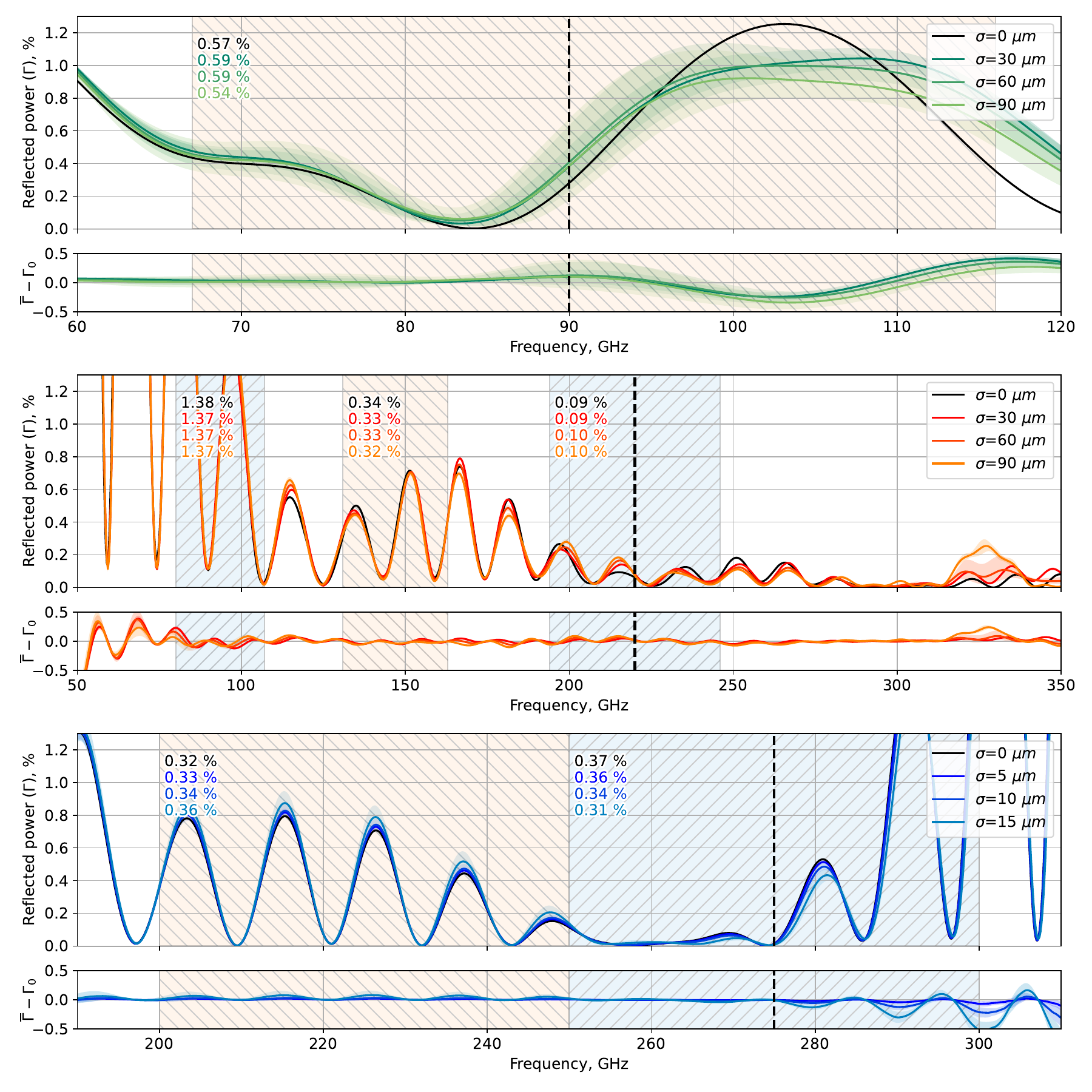}
    \caption{Reflected power as a function of frequency. Solid lines show the mean result over all iterations, transparent regions show the standard deviation around that mean. The numbers on the top left of each band show the mean value in band. The bottom panel shows the difference between the mean in randomized geometries and an ideal case with regular unit cells.}
    \label{fig:ref}
\end{figure}

The fraction of power reflected off of the ARC slab presented in Figure \ref{fig:ref} was calculated using the S11 data from the input waveguide port. The black curves in each plot correspond to the case without any random variations, virtually identical to the result that would be obtained in a unit cell simulation. The colored curves show the reflection observed with the Gaussian variation in the feature height, with the narrow plots showing the difference of these cases from the ideal case. As can be seen, the variation in reflection generally does not exceed 0.5\% for the investigated variation of geometry, with it staying below 0.1\% for most of the frequency bands of interest. Notably, while the random variation in the ARC geometry does make it less reflective when averaged across a band (see the averaged percentages on the top left of each band in Figure \ref{fig:ref}), at some frequencies the sample does become more transparent. This will be important for the later evaluation of the impact of the random geometry on power transmitted.

\begin{figure}
    \centering
    \includegraphics[width=1\linewidth]{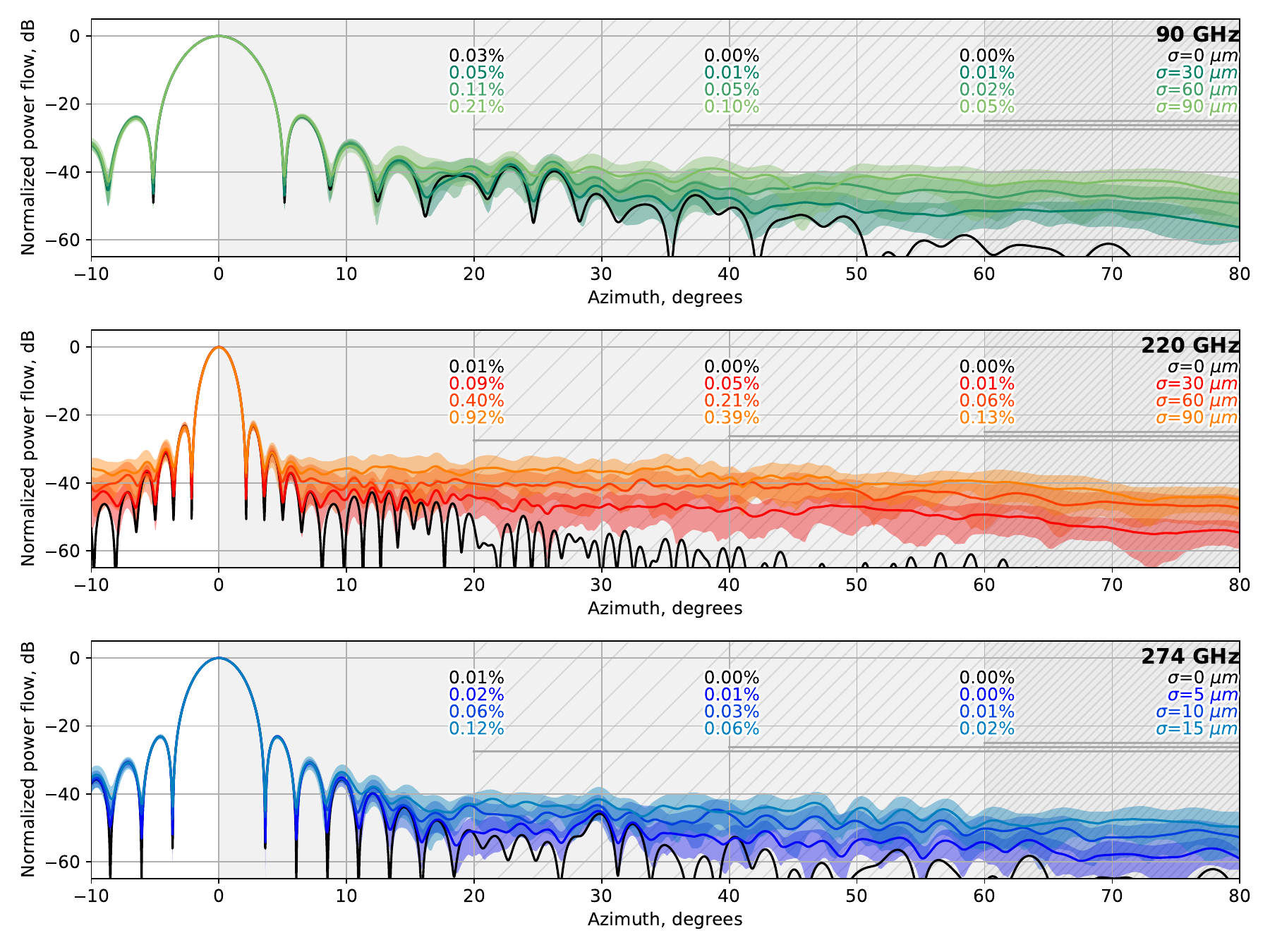}
    \caption{Far field power vs azimuthal angle. The middle frequency of the highest band is shown. Solid lines show the mean; transparent regions show the standard deviation. The colored percentages indicate the ratio of the integrated fraction of power scattered to angles above 20, 40 and 60 degrees (hatched regions) to the far field power integrated from 0 to 80 degrees.}
    \label{fig:sca}
\end{figure}

\begin{figure}
    \centering
    \includegraphics[width=1\linewidth]{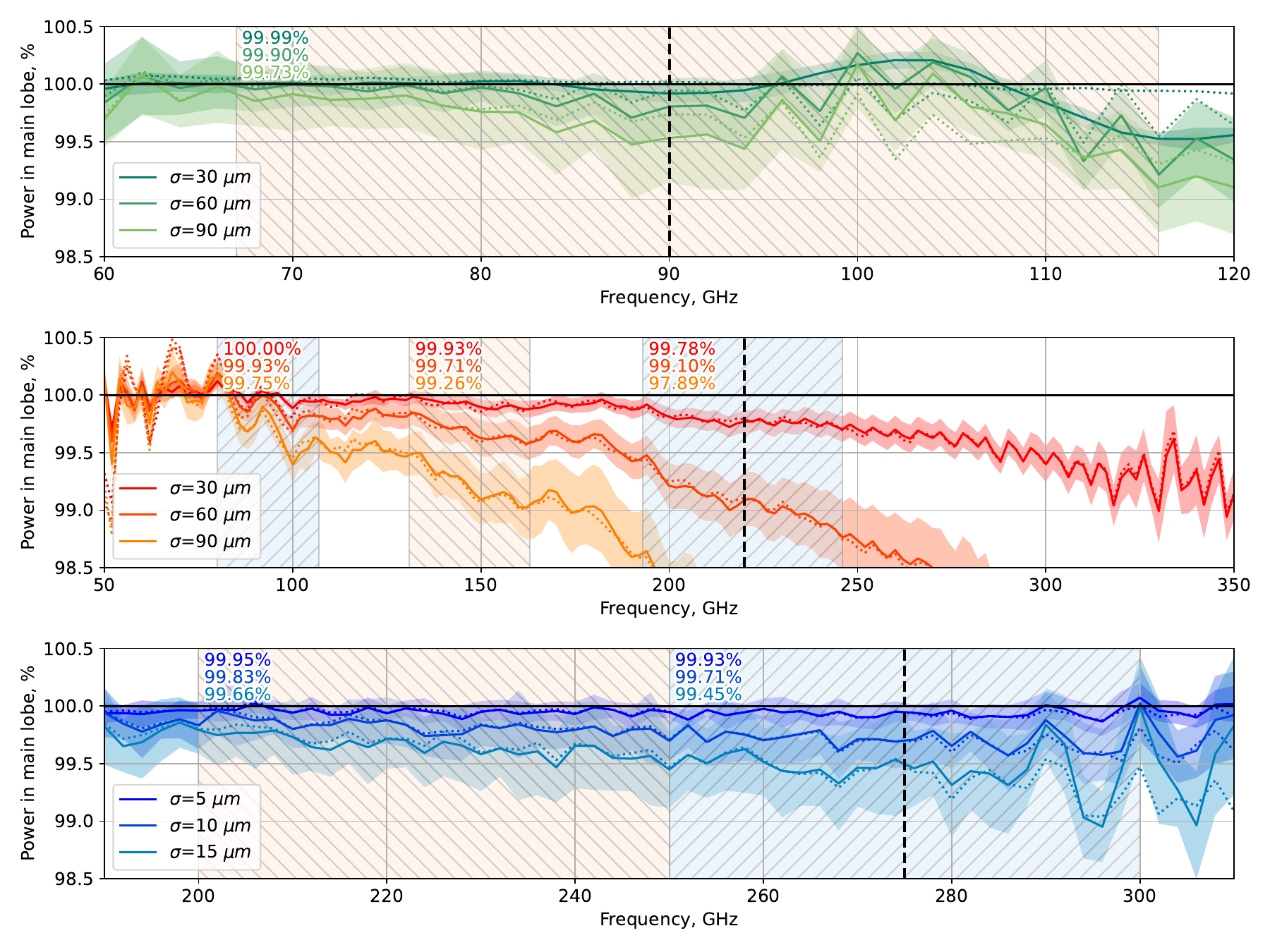}
    \caption{Transmitted power vs frequency. The quantity shown is the integrated power in the main lobe (defined as the range of angles between the first nulls) as a percentage of main lobe power for nonrandomized case. Solid lines show the mean; transparent regions show the standard deviation. The numbers on the top left of each band show the mean value in band. Dotted lines show the mean value normalized by adding the difference in reflected power to account for the change in ARC’s transparency.}
    \label{fig:tra}
\end{figure}

The amount of power scattered from the ARC slab to the far field is shown in Figure \ref{fig:sca}. Each subfigure shows the scattered power for the mean frequency of the highest band considered for the corresponding geometry (shown in a dashed black vertical line in Fig.\ \ref{fig:ref} and \ref{fig:tra}) The traces are normalized to the peak of the non-random case, shown in black. While the simulation does provide the far field across the entire sphere, due to the one-dimensional nature of the utilized unit cell shape perturbation, we have decided that a slice along the azimuth is the most informative. For the same reason, the integrated scattered power was calculated as $\int_{\theta_0}^{\theta_1}P(\theta)d\theta$ rather than $\int_{\theta_0}^{\theta_1}P(\theta)\sin{\theta}d\theta$. 

The latter assumes a uniform distribution of the scattered power along $\phi$, which would be the case if we were to simulate a completely random, non-periodic surface or a proper finite-sized 2D slab of the ARC sample, but does not describe the case with the geometry varying along a single period. The difference in the latter case case being that the scattering happens along a single preferred plane rather than uniformly across $\phi$. 
This integrated scattered power was then used to estimate the amount of power scattered to the large angles by integrating the far field starting from $\theta_0 = $ 20, 40 or 60 degrees and up to $\theta_1 = $ 80 degrees and then dividing the resulting value by the integral from 0 to 80 degrees. The reason 80 degrees was chosen for the upper value of integration rather than 90 is so that we can more easily compare the simulation results to their eventual experimental verification using a holographic robot-arm measurement \cite{Balafendiev2024}. 
The resulting values are shown in Figure \ref{fig:sca} as black (for the regular case) or colored (for the case with random perturbations) percentages, with each column corresponding to a different starting angle in the integration bounds and each row (each hue) to a different $\sigma$ value. 

As can be seen, the perturbation of the ARC geometry on these scales leads to about \SI{-40}{\decibel} ``noise floor" across all angles. However, even in the most extreme case (Nylon pyramids with $\sigma= \SI{90}{\micro\meter}$) the amount of power scattered to the angles above 20~degrees does not exceed 1\%. This noise floor can be contextualized using the near field of the ARC: the near field is tied to the far field via a Fourier transform. Once random variations of geometry are introduced, random shifts can also be observed in the magnitude and phase of the near field. To some extent, these random shifts can be thought of as additive Gaussian white noise in the magnitude and phase profile of a Gaussian beam. The Fourier transform of such Gaussian white noise with a deviation of $\sigma$ is a constant value, $\sigma^2$, which is what we observe being added to the Fourier image of the Gaussian beam in the far field.

Finally, Figure \ref{fig:tra} shows the amount of power transmitted in the main lobe as a function of frequency and the deviation of geometry. Here, the angular range between the two lobes nearest to the main lobe was chosen as the extent of the integration. The transmitted power from a perturbed geometry is normalized to the power from a perfectly regular one. Just like in Figure \ref{fig:ref}, the percentages on the top left of each band show the value averaged across the band. The solid lines show the initial result of that normalization. The reason it sometimes crosses over the 100\% line is because of how the ARC becomes more transparent at these frequencies (compare with Fig.\ \ref{fig:ref}). If the reflection data is taken into account by subtracting the normalized difference in reflection from the normalized difference in transmission, a frequency dependence that exhibits a stronger downward trend can be obtained. This result is plotted in Figure \ref{fig:tra} with dotted lines.

\begin{figure}
    \centering
    \includegraphics[width=0.85\linewidth]{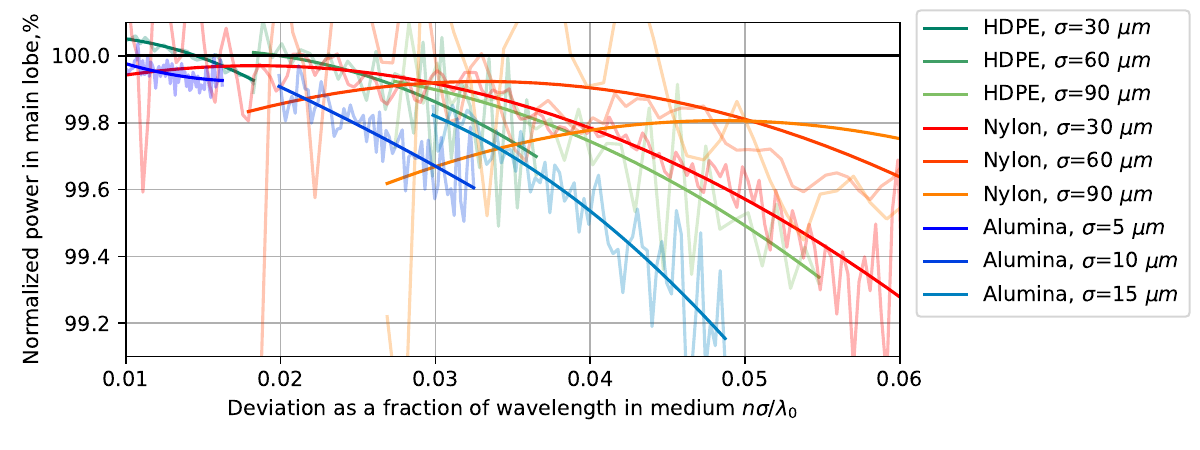}
    \caption{Power integrated over the main lobe, with the impact of variations of reflection factored in (see dotted lines in Fig.\ \ref{fig:tra}). The solid lines are quadratic polynomials fit over the data (transparent lines). }
    \label{fig:tot}
\end{figure}

The same reflection-corrected data is plotted again in Figure \ref{fig:tot}, this time as a function of the deviation in geometry normalized to the wavelength in medium. The raw data is plotted with transparent lines, while the solid lines show quadratic polinomial fits over the raw data. The reason for choosing a quadratic fit specifically was, in broad terms, the $\sigma^2$ constant value expected from the Fourier image of the perturbed near field. In fact, the same behavior of the transmitted power falling as $\sigma^2$ can be expected from a similar result presented in a paper by Ruze\cite{Ruze1966}. As can be seen from the figure, the raw data does generally follow a quadratic curve fairly well. While normalizing the $\sigma$ value to the wavelength is straightforward, it can be speculated that with a more precise characterization of how much do these deviations affect the magnitude and the phase for each geometry all of the obtained data can be brought to a single quadratic curve. This would, however, require a more in-depth characterization of how various variations of the geometry affect transmission in a unit cell case, similarly to how it was done in Datta et al.\cite{Datta2013}. The overall conclusion that can be drawn from this figure is the following: generally, variations of the unit cell geometry within 5\% of the wavelength in the material at a given frequency will lead to less that 1\% of loss of power in transmission.

\section{CONCLUSIONS}
The variation in the geometry of the ARCs leads to an increased scattering to all angles and a decreased power in the main lobe of the beam. By making a link between variations in geometry and variations in phase, this can be directly tied to the analogous conclusion reached in works of Ruze and Lamb\cite{Ruze1966,Lamb1996a}. 
The approach presented in this work can and is planned to be applied to other, less random variations in geometry, such as curved lens surfaces or linear errors in depth due saws being worn down when machining hard materials. It can also be used to estimate the effect of various local or nonlocal variations in ARC on polarization. We plan to expand on this by investigating these aspects and experimentally verifying the results presented here in a future publication.

\acknowledgments % equivalent to \section*{ACKNOWLEDGMENTS}       
The authors thank Xiaodong Ren and Pavel Belov for helpful discussions. Funded in part by the European Union (ERC, CMBeam, 101040169). JEG gratefully acknowledges support from the University of Iceland Research Fund and the Icelandic Research Fund (Grant number: 2410656-051).

% References
\bibliography{report} % bibliography data in report.bib

@article{Datta2013,
  title = {Large-Aperture Wide-Bandwidth Antireflection-Coated Silicon Lenses for Millimeter Wavelengths},
  author = {Datta, R. and Munson, C. D. and Niemack, M. D. and McMahon, J. J. and Britton, J. and Wollack, E. J. and Beall, J. and Devlin, M. J. and Fowler, J. and Gallardo, P. and Hubmayr, J. and Irwin, K. and Newburgh, L. and Nibarger, J. P. and Page, L. and Quijada, M. A. and Schmitt, B. L. and Staggs, S. T. and Thornton, R. and Zhang, L.},
  date = {2013-12-20},
  journaltitle = {Applied Optics},
  shortjournal = {Appl. Opt.},
  volume = {52},
  number = {36},
  pages = {8747},
  issn = {1559-128X, 2155-3165},
  doi = {10.1364/AO.52.008747},
  url = {https://opg.optica.org/abstract.cfm?URI=ao-52-36-8747},
  urldate = {2026-07-24},
  langid = {english}
}

@article{Golec:22,
  title = {Simons {{Observatory}}: Broadband Metamaterial Antireflection Cuttings Forlarge-Aperture Alumina Optics},
  author = {Golec, Joseph E. and Sutariya, Shreya and Jackson, Rebecca and Zimmerman, Jerry and Dicker, Simon R. and Iuliano, Jeffrey and McMahon, Jeff and Puglisi, Giuseppe and Tucker, Carole and Wollack, Edward J.},
  date = {2022-10},
  journaltitle = {Applied Optics},
  shortjournal = {Appl. Opt.},
  volume = {61},
  number = {30},
  pages = {8904--8911},
  publisher = {Optica Publishing Group},
  doi = {10.1364/AO.472459},
  url = {https://opg.optica.org/ao/abstract.cfm?URI=ao-61-30-8904}
}

@article{Lamb1996,
  title = {Miscellaneous Data on Materials for Millimetre and Submillimetre Optics},
  author = {Lamb, James W.},
  date = {1996-12},
  journaltitle = {International Journal of Infrared and Millimeter Waves},
  shortjournal = {Int J Infrared Milli Waves},
  volume = {17},
  number = {12},
  pages = {1997--2034},
  issn = {0195-9271, 1572-9559},
  doi = {10.1007/BF02069487},
  url = {http://link.springer.com/10.1007/BF02069487},
  urldate = {2026-07-24},
  langid = {english}
}

@article{Lamb1996a,
  title = {Cross-Polarisation and Astigmatism in Matching Grooves},
  author = {Lamb, James W.},
  date = {1996-12},
  journaltitle = {International Journal of Infrared and Millimeter Waves},
  shortjournal = {Int J Infrared Milli Waves},
  volume = {17},
  number = {12},
  pages = {2159--2165},
  issn = {0195-9271, 1572-9559},
  doi = {10.1007/BF02069491},
  url = {http://link.springer.com/10.1007/BF02069491},
  urldate = {2026-07-24},
  langid = {english}
}

@article{Nitta2018,
  title = {Design, {{Fabrication}} and {{Measurement}} of {{Pyramid-Type Antireflective Structures}} on {{Columnar Crystal Silicon Lens}} for {{Millimeter-Wave Astronomy}}},
  author = {Nitta, T. and Sekimoto, Y. and Hasebe, T. and Noda, K. and Sekiguchi, S. and Nagai, M. and Hattori, S. and Murayama, Y. and Matsuo, H. and Dominjon, A. and Shan, W. and Naruse, M. and Kuno, N. and Nakai, N.},
  date = {2018-12},
  journaltitle = {Journal of Low Temperature Physics},
  shortjournal = {J Low Temp Phys},
  volume = {193},
  number = {5--6},
  pages = {976--983},
  issn = {0022-2291, 1573-7357},
  doi = {10.1007/s10909-018-2047-4},
  url = {http://link.springer.com/10.1007/s10909-018-2047-4},
  urldate = {2026-07-24},
  langid = {english}
}

@article{Ruze1966,
  title = {Antenna Tolerance Theory -- A Review},
  author = {Ruze, J.},
  date = {1966},
  journaltitle = {Proceedings of the IEEE},
  shortjournal = {Proc. IEEE},
  volume = {54},
  number = {4},
  pages = {633--640},
  issn = {0018-9219},
  doi = {10.1109/PROC.1966.4784},
  url = {http://ieeexplore.ieee.org/document/1446714/},
  urldate = {2026-07-24}
}

@article{Tapia2018,
  title = {Systematic Study of the Cross Polarization Introduced by Broadband Antireflection Layers at Microwave Frequencies},
  author = {Tapia, Valeria and Rodríguez, Rafael and Reyes, Nicolás and Patricio Mena, F. and Yagoubov, Pavel and Cuttaia, Francesco and Bronfman, Leonardo},
  date = {2018-11-01},
  journaltitle = {Applied Optics},
  shortjournal = {Appl. Opt.},
  volume = {57},
  number = {31},
  pages = {9223},
  issn = {1559-128X, 2155-3165},
  doi = {10.1364/AO.57.009223},
  url = {https://opg.optica.org/abstract.cfm?URI=ao-57-31-9223},
  urldate = {2026-07-24},
  langid = {english}
}

@article{Xue2024,
  title = {Complex {{Permittivity Characterization}} of {{Low-Loss Dielectric Slabs}} at {{Sub-THz}}},
  author = {Xue, Bing and Haneda, Katsuyuki and Icheln, Clemens and Ala-Laurinaho, Juha and Tuomela, Juha},
  date = {2024},
  journaltitle = {IEEE Access},
  shortjournal = {IEEE Access},
  volume = {12},
  pages = {150693--150701},
  issn = {2169-3536},
  doi = {10.1109/ACCESS.2024.3479908},
  url = {https://ieeexplore.ieee.org/document/10716426/},
  urldate = {2026-07-24}
}

@inproceedings{Balafendiev2024,
  title = {Vector Beam Mapping at Millimeter Wavelengths Using a Robot Arm},
  booktitle = {Millimeter, {{Submillimeter}}, and {{Far-Infrared Detectors}} and {{Instrumentation}} for {{Astronomy XII}}},
  author = {Balafendiev, Rustam and Gascard, Thomas and Gudmundsson, Jon E.},
  editor = {Zmuidzinas, Jonas and Gao, Jian-Rong},
  date = {2024-08-16},
  pages = {134},
  publisher = {SPIE},
  location = {Yokohama, Japan},
  doi = {10.1117/12.3020315},
  url = {https://www.spiedigitallibrary.org/conference-proceedings-of-spie/13102/3020315/Vector-beam-mapping-at-millimeter-wavelengths-using-a-robot-arm/10.1117/12.3020315.full},
  urldate = {2026-07-29},
  eventtitle = {Millimeter, {{Submillimeter}}, and {{Far-Infrared Detectors}} and {{Instrumentation}} for {{Astronomy XII}}},
  isbn = {978-1-5106-7527-8 978-1-5106-7528-5}
}
\bibliographystyle{spiebib} % makes bibtex use spiebib.bst

\end{document}